\documentclass[sn-mathphys,Numbered]{sn-jnl}

\usepackage{graphicx}%
\usepackage{multirow}%
\usepackage{amsmath,amssymb,amsfonts}%
\usepackage{amsthm}%
\usepackage{mathrsfs}%
\usepackage[title]{appendix}%
\usepackage{xcolor}%
\usepackage{textcomp}%
\usepackage{manyfoot}%
\usepackage{booktabs}%
\usepackage{algorithm}%
\usepackage{algorithmicx}%
\usepackage{algpseudocode}%
\usepackage{listings}%
\usepackage{amssymb}
\usepackage{array}
\usepackage{graphicx}
\usepackage{float}
\usepackage{color,soul}
\usepackage{amsmath}
\usepackage{amsthm}
\usepackage{amsopn}

\def\be{\begin{equation}}
	\def\ee{\end{equation}}
\def\bea{\begin{eqnarray}}
	\def\eea{\end{eqnarray}}

\theoremstyle{thmstyleone}%
\theoremstyle{thmstyletwo}%

\theoremstyle{thmstylethree}%

\begin{document}

\title[Article Title]{Efficiency of Higher Dimensional Black Holes as Particle Accelerators}


\author[1]{\fnm{Fatemeh} \sur{Behdadkia}}\email{f.behdadkia.fb@gmail.com}

\author*[1]{\fnm{Behrouz} \sur{Mirza}}\email{b.mirza@iut.ac.ir}

\author*[1]{\fnm{Masoumeh} \sur{Tavakoli}}\email{tavakoli.phy@gmail.com}


\affil[1]{\orgdiv{Department of Physics}, \orgname{Isfahan University of Technology}, \orgaddress{ \postcode{84156-83111}, \state{Isfahan}, \country{Iran}}}



\abstract{The center-of-mass energy of two colliding particles could be arbitrarily high in the vicinity of event horizons of the extremal Myers-Perry black holes if the angular momentum of colliding particles is fine-tuned to the critical values. We investigate the maximum efficiency of two colliding particles in four and six dimensions. The efficiency of collision for two particles near the four-dimensional Kerr black holes is 130\%. We show that the efficiency increases to 145\% for collision in six dimensions. We also show that the region for the polar angle in which the particle can reach the high energy is larger when the dimension of space-time increases.}




\maketitle

\section{Introduction}\label{sec:intro}

It is known that extremal Kerr black hole can accelerate particles in the vicinity of their event horizons. This effect which is known as Banados, Silk, West (BSW) effect could happen when an angular momentum of one of particles that participates in the collision is fine-tuned to a critical value  \cite{09090169}.

In different studies, it was shown that various rotating space-times can accelearte particles to high energy \cite{10061056,14115778,14105588,150308553,160705063,12024159,11106274,09113363,10015463}. 
The BSW effect may also occur in the vicinity of non-extremal Kerr black holes. The particles can reach high energy near the event horizon of non-extremal black holes through multiple scattering \cite{10040913}. The center-of-mass (CM) energy of two colliding general geodesic particles for non-extremal Kerr black holes were  calculated in \cite{11023316}. Also in this study, a polar region in which high energy collision is allowed was found.
A collision of particles that move on innermost stable circular orbit (ISCO) was studied in \cite{10100962}.
It was shown that the center-of-mass energy of two particles which collide on  the ISCO is also arbitrary high. The CM energy for Compton scattering, pair annihilation and elastic collision in the vicinity of Kerr black holes were studied in  \cite{12057088}. It was shown than Campton scattering is the most efficient one among these processes.

Using the efficiency of collision near the Kerr black hole an upper limit was suggested for the CM energy which extract from extremel kerr black holes. It was shown that the peak efficiency for colliding particle near the extremal Kerr black hole is 130\% . Peak Efficincy for Compton scattering and pair annihilation were also obtained \cite{14106446,12054350}.

Also, in \cite{13105716,170809576} Myers--Perry (MP) black holes were analayzed. It was argued that extremal MP black holes can accelerate particles near their event horizions when the angular momentum of paritcle  is fine tuned to the critial value momentum. Also collision of particles near five dimensional black string was studied in \cite{13111751}. Dynamics of particles  near the higher dimension black holes were investigated in \cite{160703507}. Some  general arguments for describing accelerating particle can be found in  \cite{10110167,10073678,14094024}. In this paper, for the first time we calculate the efficiency of collion for six dimensional rotating black holes.

The outline of this paper is as follow. In Section II, we shortly  review the MP black holes and consider particular case  in which MP black holes have two distinct spin parameters. In Section III, we briefly explain a method to obtain geodesic equations in even  dimensions for MP black holes.  We also introduce effective potentials for  even dimensions by using radial geodesic equations. In Section IV,   for different even dimensions, the CM energy and effective potential diagrams are depicted. The polar angle region in which particle  can reach high energy is obatianed. In Section V,  the maximum efficiency of two colliding particles is studied for higher dimensional rotating black holes.
\section{The Myers--Perry black holes}\label{sec2}
In this Section, we consider the higher dimensional Myers--Perry black holes for D= 6, 8, 10,... \cite{11111903}. 
We assume that all spins are equal in a $D = 2n + 2$ dimensional space time.

To begin with, the metric of the even  dimensional ($D\geq 6$) MP black holes can be written as below:
\begin{eqnarray}\label{a5}
	ds^2 &=-dt^2+\frac{2Mr}{\Pi F}(dt-\sum _{i=1}^{n}a_{i}\mu _{i}^{2}d\phi    _{i})^2+\frac{\Pi F}{\Pi -2Mr}dr^2\nonumber\\
	& +\sum_{i=1}^n(r^2+a^2_i)(d\mu _i ^2+\mu _i^2d\phi _i^2) +r^2d\alpha ^2,
\end{eqnarray}
where  $M$ and $a_i$ represent mass and spin parameters, respectively.
Also  $\Pi$ and $F$ are given by Eqs.(\ref{a2}) and (\ref{a3}), respectively.
\begin{eqnarray}
	&\Pi=\prod ^{n}_{i=1}(r^2+a_i^2),\label{a2}\\  
	&F=1-\sum _{i=1}^{n}\frac{a_i^2\mu ^2_i}{r^2+a^2_i},\label{a3}\\
	&\sum _{i=1}^{n}\mu _i^2+\alpha ^2=1\label{a6}.
\end{eqnarray}

For simplicity, we assume the case that spin parameters are equal. 

According to the division of spin parameters, $\mu _i$ can be decomposed as follows:
\begin{eqnarray}\label{a8}
	\mu _i=\lambda _i \sin \theta ,\hspace{2cm}i=1, ..., n.
\end{eqnarray} 
where n= (D-2)/2 and  $\sum_{i=1}^{n} \lambda _i ^2 =1$.

\begin{figure}[htbp!]
	\centering
	\includegraphics[scale=0.2]{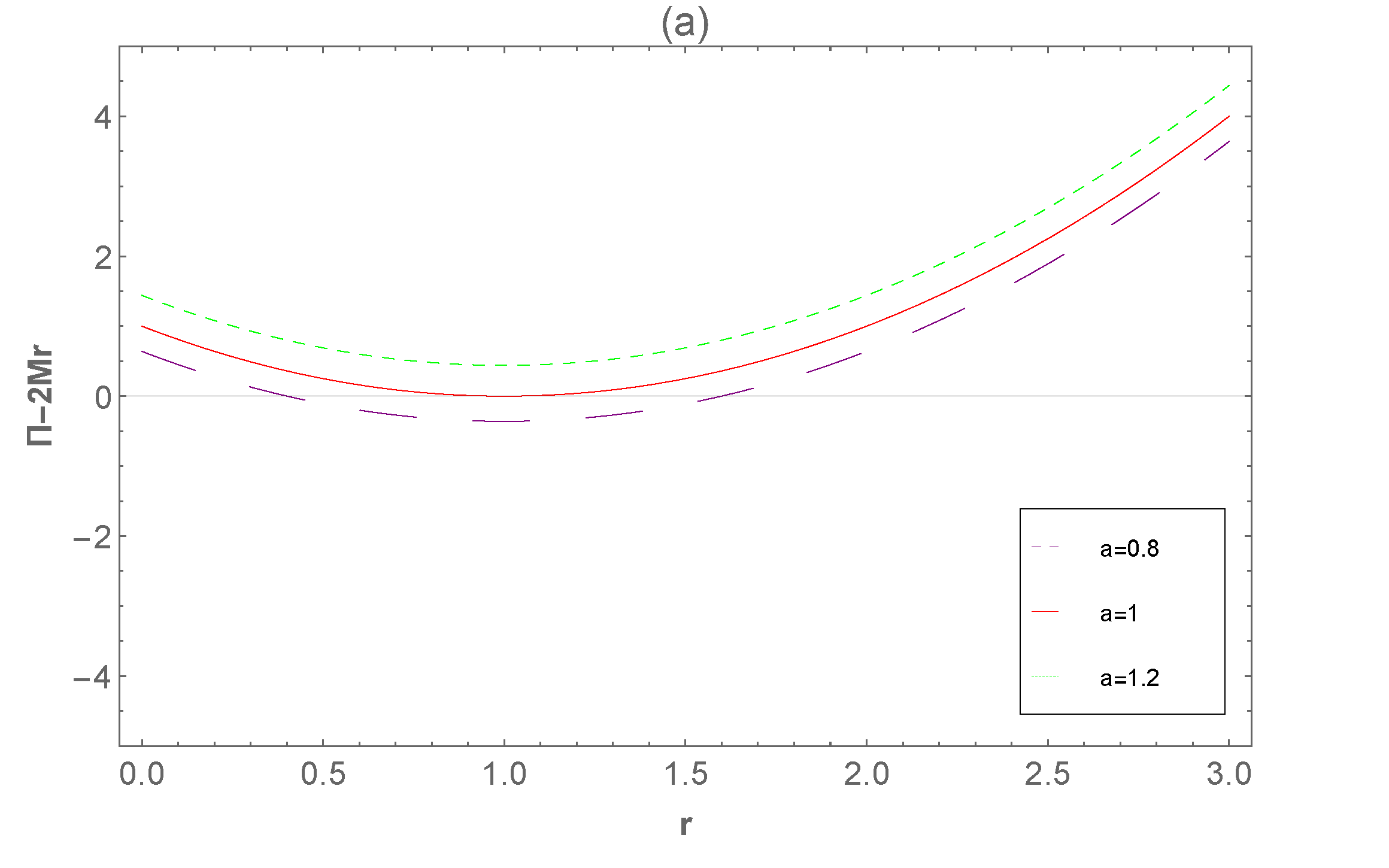}
	\hspace{0.5cm}
	\includegraphics[scale=0.2]{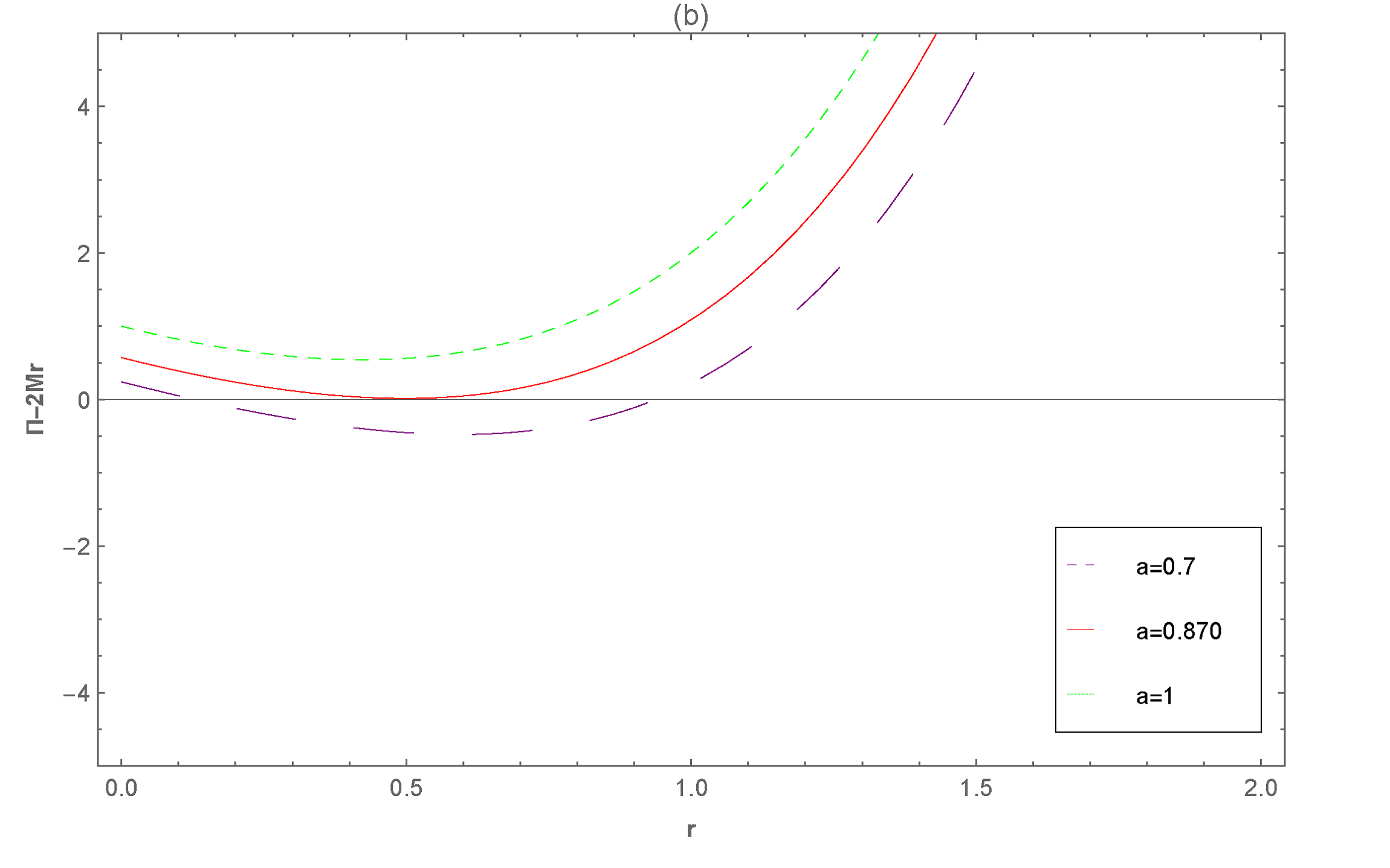}
	\caption{\label{evenhorizon} In these plots, behaviour of $\Delta _{even}$ in Eq.(\ref{a11}) versus $r$ are depicted in dimensions 4 and 6 for different value of spin parameters $a$. Extremal cases are plotted in solid red lines. (a) Four-dimensions with n=1. (b) Six-dimensions with n=2.  In all cases, we put $M=1$ in which $M$ represent mass of black holes.}
\end{figure}

With equal  spin parameters in even dimensions, we can define the following functions:
\begin{eqnarray}
	&\Pi _{even}=(r^2+a^2)^n,    
	\label{a9}\\
	&Z_{even}=r^2(r^2+a^2),\label{a10}\\
	&\Delta _{even}=\Pi _{even}-2Mr.\label{a11}
\end{eqnarray}  
$\Delta_{even}$ are equal to zero on the horizons of the MP black holes.
The event horizon is the largest root of  $\Delta _{even}$.  
For  simplicity,  we can consider the extremal case where the event horizon is the double root. 
The double root of an equation is a common value that make  $\Delta _{ even}$ and its first derivative equal to zero. 
In the extremal case, the event horizon for even dimensions are located at 
\begin{eqnarray}
	r_{+}^{even}=\sqrt{\frac{a}{2n-1}}.\label{a13}
\end{eqnarray}  
The event horizon of non-extremal and extremal black holes are depicted for different values of even dimensions in Fig \ref{evenhorizon}.

\section{Geodesic equations of MP black holes}\label{sec3}
In this Section, geodesic equations of MP black holes in even dimensions are briefly reviewed \cite{0407030}. To obtain geodesic equations, we use Hamilton--Jacobi equation as follows:
\begin{eqnarray}
	&\frac{\partial S}{\partial \lambda}+H=0,\label{b1}\\
	&H=\frac{1}{2}\sum _{\mu \nu}g^{\mu \nu}p_{\mu}p_{\nu},\label{b2}\\
	&p_{\alpha}=\frac{\partial S}{\partial x^\alpha},\label{b3}
\end{eqnarray}
where $p_\alpha$, $S = S(\lambda, x^\mu)$, $g^{\mu \nu}$ and $H$ are conjugated momenta, action function, contravariant components of the MP metric, and Hamiltonian, respectively.
Due to cyclic coordinates $t$ and $\phi _i$, their conjugated momenta are constant. These constants are energy and angular momenta of particles, which are accelerated near horizon of the MP black holes. Energy and angular momenta of particles are shown with $E$ and $\Phi _i$  respectively. Therefore, with respect to these cyclic coordinates and their conjugated momenta, the action function $S$ in even dimensions can be written as follows:
\begin{eqnarray}\label{eq14}
	S=\frac{1}{2}m^2 \lambda -Et+\sum _{i=1}^{n}\Phi _i \phi _i+S_r(r)+S_{\theta}(\theta)
	+\sum _{i=1}^{n-1}S_{\alpha _i}(\alpha _i).
\end{eqnarray}
where, $m$ and $\lambda$ are the mass of particle and affine parameter respectively. 

By substituting  Eqs.(\ref{b3}), (\ref{eq14}) and contravaraint components of metric  Eq.(\ref{a5}) in Hamilton--Jacobi equation, Eq.(\ref{b1}), we get a complicated equation. Using separation of constant $J_1^2$ which is given by
\begin{eqnarray}
	&J_1^2=\sum _{i=1}^{n}[\frac{\Phi _i^2}{\lambda _i ^2}+\frac{1}{\prod _{k=1}^{i-1}\sin ^2 \alpha _k}\big(\frac{dS_{\alpha  _i}}{d\alpha _i}\big)^2],\label{b5}
\end{eqnarray}
we may rewrite the equation in the form  that each  sides of it depend on $r$ and $\theta$. We put both sides of equation equal to a constant $K$.

\begin{eqnarray}
	&K= (E^2-m^2)r^2+\frac{2MZ_{even}}{r\Delta_{even}}P(r)^2+\frac{a^2 J_1^2}{r^2+a^2}-\frac{\Delta(r^2+a^2)}{\Pi_{even}}(\frac{dS_r}{dr})^2\label{b701}
	\nonumber\\
	& K=\frac{J_1^2}{\sin ^2\theta}+\big(\frac{dS_{\theta}} {d\theta}\big)^2+(m^2-E^2)a^2\cos ^2\theta.\label{b702}
\end{eqnarray}  
where $P(r)$ is defined as below:
\begin{eqnarray}\label{b92}
	P(r)=E-\frac{a}{r^2+a^2}\sum _{i=1}^{n}\Phi _i,\label{b29}
\end{eqnarray}

To obtain $r$ and $\theta$ equations of motion, we should integrate Eqs. (\ref{b701}) and (\ref{b702})
\begin{eqnarray}\label{b8}
	S_{\theta}= \sigma _ {\theta} 
	\int d\theta \sqrt{\Theta(\theta)},\hspace{2cm} S_{r}=\sigma _{r}\int dr\sqrt{R(r)},
\end{eqnarray}
where for inward and outward  particle motion to black hole, we have $\sigma _{r,\theta}=+1$ and $\sigma _{r,\theta}=-1$, respectively. Also, $\Theta (\theta)$ and $R(r)$ are given by
\begin{eqnarray}
	&\frac{r^2+a^2}{\Pi _{even}} R(r)=2Mr(r^2+a^2)P(r)^2-\Delta _{even}[m^2r^2+K-\frac{(E(r^2+a^2)-  aJ_1^2)^2}{r^2+a^2}],\label{b28}\\
	&\Theta(\theta)=K-m^2\cos ^2\theta-\frac{1}{\sin ^2\theta}(aE\sin ^2\theta -J_1)^2\label{b30}.	
\end{eqnarray}

To full separate Hamilton--Jacobi equation, we  separate Eqs.(\ref{b5}) ) to two  parts  as follows:
\begin{eqnarray}\label{b9}                 
	&\Big(\frac{dS_{\alpha _k}}{d\alpha _k}\Big) ^2=J^2_k-\frac{J^2_{k+1}}{\sin ^2\alpha _k}
	-\frac{\Phi ^2_{n-k+1}}{\cos ^2\alpha _k},\nonumber\\
	&\Big(\frac{dS_{\alpha _{n-1}}}{d\alpha _{n-1}}\Big) ^2=J^2_{n-1}-\frac{\Phi ^ 2_{1}}{\sin ^2\alpha             
		_{n-1}}-\frac{\Phi ^2_{2}}{\cos ^2\alpha _{n-1}},\nonumber\\
\end{eqnarray}
where $J_k^2$  are separation constants.
We should integrate all equations in (\ref{b9}) to obtain equations of motion for $\alpha _k$.
\begin{eqnarray}\label{b10}
	S_{\alpha}&=\sigma_{\alpha}\int^{\alpha _k}d\alpha _k \sqrt{A_k(\alpha _k)},     
\end{eqnarray}
where $\sigma _{\alpha} =\pm 1$  correspond to ingoing and outgoing direction, respectively. Also, $A_k(\alpha _k)$ are defined as follows:
\begin{eqnarray}
	&A_k &=J^2_k-\frac{J^2_{k+1}}{\sin ^2\alpha _k}-\frac{\Phi ^2_{q-k+1}}{\cos ^2\alpha _k},\hspace{2cm}k=1,...,n-2,\nonumber\\
	&A_{n-1}&=J^2_{n-1}-\frac{\Phi^ 2_{1}}{\sin ^2\alpha _{n-1}}-\frac{\Phi ^2_{2}}{\cos ^2\alpha _{n-1}},\label{b110}\\
\end{eqnarray}
Using $\frac{dx^{\alpha}}{d\lambda}=\sum _{\beta}g^{\alpha \beta}p_{\beta}$, we write geodesic equations of black holes in even dimensions as follows: 
\begin{eqnarray}
	& \rho ^2 \frac{dt}{d\lambda}\equiv \rho ^2 \dot{t}=E\rho ^2+\frac{2MZ_{even}}{r\Delta _{even}}P(r),\label{b23}\\
	&\rho ^2 \frac{dr}{d\lambda} \equiv \rho ^2\dot{r}=\sigma _{r}\frac{r^2+a^2}{\Pi _{even}}\sqrt{R(r)},\label{b24}\\
	&\rho ^2 \frac{d\theta}{d\lambda} \equiv \rho ^2\dot{\theta}=\sigma _{\theta}\sqrt{\Theta(\theta)},\label{b25}\\
	&\rho ^2 \frac{d \phi _i}{d\lambda} \equiv \rho ^2 \dot{\phi _i }=\frac{\rho ^2}{\lambda ^2_i \sin ^2\theta(r^2+a^2)}\Phi _i+\frac{2aMr}{\Delta _{even}}P(r),\label{b26}\\
	&(r^2+a^2)\frac{d\alpha _k}{d\lambda}\equiv \sigma _{\alpha _k}(r^2+a^2)\dot{\alpha _k}=\frac{\sqrt{A _k}}{\sin ^2\theta \prod _{i=1}^{k-1}\sin ^2\alpha _i}.\label{b27}
\end{eqnarray}
in these equations $E$ and $\Phi _i$ are energy and angular momenta of particle. Also $R(r)$, $\Theta(\theta)$,  $P(r)$, $A(\alpha _k)$  are defined in Eqs.(\ref{b28}), (\ref{b30}), (\ref{b92}), and  (\ref{b110}), respectively. In these equations, we define $\rho ^2= r^2+a^2 \cos^2 \theta $ for even dimensions. One may use the $r$ geodesic in Eq.(\ref{b24}) and rewrite it as $\frac{1}{2}\dot{r}^2+V_{eff}(r,\theta)=0$. Therefore, the effective potential for even dimensions is  as below:
\begin{eqnarray}
	&V_{eff}(r,\theta)=\frac{V(r)}{2(r^2+a^2)^{n-1}\rho^4},\label{b31}\\
	&V(r)=2Mr(r^2+a^2)P(r)^2-\Delta_{even}[m^2r^2+K-\frac{(E(r^2+a^2)-aJ_1^2)^2}{r^2+a^2}].\label{b32}
\end{eqnarray}

\section{The CM energy of two colliding particles in the vicinity of MP black holes}\label{sec4}
In this section, we consider MP black holes as particles accelerators and calculate the CM energy of two particles which collide in the vicinity of extremal black holes. We assume that rest mass and the four momentum of two colliding particles $i$ ($i=1,2$) are  equal to $m_i$ and $p_i$, respectively. The CM energy of  two colliding particles is given by
\begin{eqnarray}\label{c1}
	E^2_{CM}=-g^{\mu \nu}(p_{1\mu} +p_{2\mu})(p_{1\nu}+p_{2\nu})=m_1^2+m_2^2-2g^{\mu\nu}p_{1\mu}p_{2\nu}.
\end{eqnarray}
where $g^{\mu\nu}$ represents the contravariant components of metric. The CM energy is scalar and it is independent of  the chosen coordinates.

The CM energy of two colliding particles in even dimensions is obtained by  inserting contravariant components of metric in  Eq.(\ref{a5}) and conjugated momenta in (\ref{b3}) in  Eq.(\ref{c1}) as below:
\begin{eqnarray}\label{c2}
	E^2_{CM}=m_1^2+m_2^2+\frac{2}{\rho^2}\frac{(r^2+a^2)^{n-1}P_{(1)}P_{(2)}-\sqrt{R_{1}(r)}\sqrt{R_{2}(r)}}{\Delta _{even}}+... .
\end{eqnarray}
In this equation, $m_{(i)}$ is mass of particle $i$. Also, $\Delta _{even}(r)$, $R_i(r)$ and $P_i(r)$ are  functions  which are defined in Eqs. (\ref{a11}), (\ref{b28}) and (\ref{b92}) for even dimensional MP black holes for particles $i (i=1, 2)$.

Now, we calculate the limit of CM energy in the vicinity of event horizon of the extremal black holes.
Some parts of this limit involves indeterminate forms and the limit can be evaluated by applying two times of L'Hospital's rule. This limit blows up near the event horizon if the angular momentum is equal to a critical value for the particle participate in collision. 

In order to find the critical value for angular momentum, we use causality of particles which move on the geodesic equations. These  particles move from the past to future, so that we can write $\frac{dt}{d\lambda}\geq0$  in Eq.(\ref{b23}), and obtain $P(r)\geq0$. Therefore by using Eq.(\ref{b92}), the critical value for the angular momentum of particle is defined as below 
\begin{eqnarray}\label{c4}
	P(r_+^{even})=E-\frac{a}{r_+^2+a^2}\sum_{i=1}^{n}\Phi _i=0
\end{eqnarray}
where $\Phi_i$ are angular momenta of particles. By determining black hole Mass $M$, spin parameter $a$, particle energy $E$, we can find value for angular momentum. Numbers of angular momentum are increasing for higher dimensions of black holes. The critical value for angular momenta of particle makes the effective potential  in Eq.(\ref{b31}) and its first derivative equal to zero at the event horizons of the extremal black holes. 
We  investigate only $V(r)$ in Eq.(\ref{b31}) instead of effective potential because the coefficient $\frac{1}{2r^4(r^2+a^2)^{2n-2}}$ in the effective potential is always positive. Near the event horizons of extremal MP black holes, $V(r)$ and its first derivative are zero because its different parts are multiple of $\Delta _{even}$, $\Delta ' _{even}$, and $P(r)$.  $\Delta _{even}$ and $\Delta ' _{even}$ are zero for extremal event horizon, and $P(r)$ is zero for particles with critical angular momenta. In other words, we can say that  radial velocity of particle is zero for critical values of angular momenta.

\begin{figure}[htbp!]
	\centering
	\includegraphics[scale=0.25]{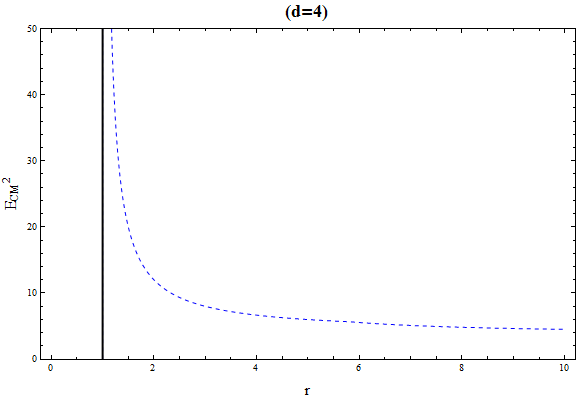}
	\hspace{0.5cm}
	\includegraphics[scale=0.25]{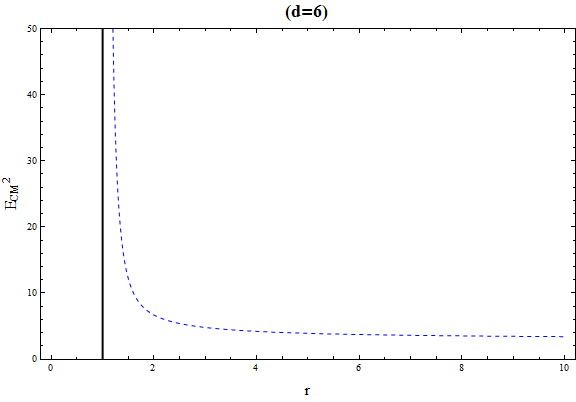}
	\hspace{0.5cm}
	\includegraphics[scale=0.25]{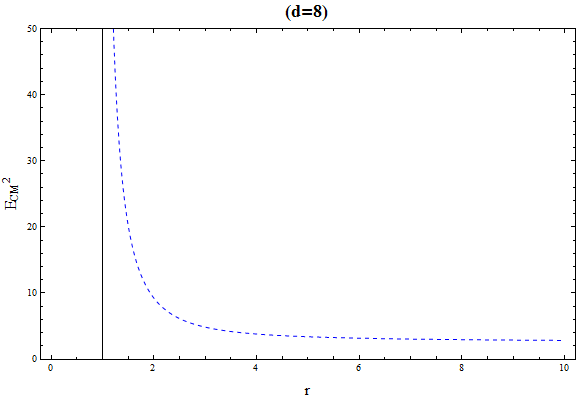}
	\caption{\label{evenecom} The CM energy of two colliding particles versus $r$ are depicted for 4, 6 and 8 dimensions. This energy blows up near the event horizon of the extremal black holes. The vertical lines show the event horizon.. For $d=4$ black hole mass ($M$) is equal to 1, and spin parameter ($a$) is equal to 1. For $d=6$, $M=8$ and $a=\sqrt{23}$. For $d=8$, $M=108$ and $a=\sqrt{5}$. }
\end{figure}

The effective potentials for critical particles are shown in Fig. {\ref{evenveff} for the extremal black holes. At the event horizon, effective potential is zero and maximum. In other words, we vary $n$ value which lead to effective potential for different dimensions. Additionally, these plots are depicted for specific values mass $M$ and spin parameter $a$ of black holes in extremal cases. maximum of effective potential.

	\begin{figure}[htbp!]
		\centering
		\includegraphics[scale=0.3]{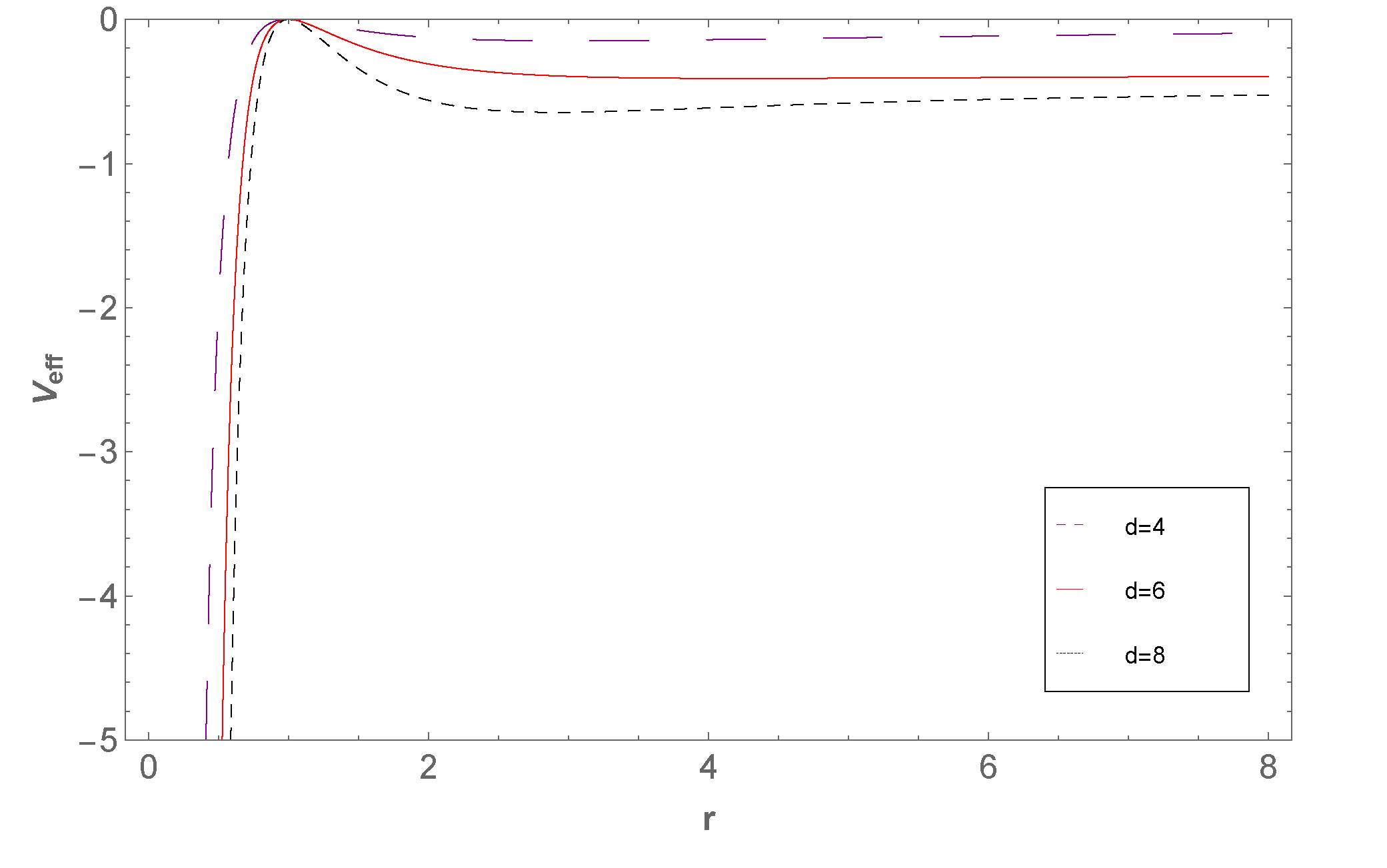}
		\caption{\label{evenveff}The effective
			potentials (Eq.(\ref{b31})) are depicted 
			for critical particles in 4, 6 and  8 dimensions.For $d=4$ black hole mass ($M$) is equal to 1, and spin parameter ($a$) is equal to 1. For $d=6$, $M=8$ and $a=\sqrt{23}$. For $d=8$, $M=108$ and $a=\sqrt{5}$.}
	\end{figure}

	The CM energy of two colliding particle for 4, 6 and 8 dimensions are depicted in Fig. \ref{evenecom}}.
We assumed that particles collide in the equatorial plane of space-time with only one non-zero angular momentum $\Phi _1$. By considering one non-zero angular momentum and using event horizon of extremal black holes for even dimensions in Eq.(\ref{a13}), we can rewrite Eq.(\ref{b5}) as below
\begin{eqnarray}\label{d3}
	J_1^2=\frac{(r_+^2+a^2)^2E^2}{a^2}=\frac{4n^2}{2n-1}a^2E^2.
\end{eqnarray}

\begin{figure}[htbp!]
	\centering
	\includegraphics[scale=0.3]{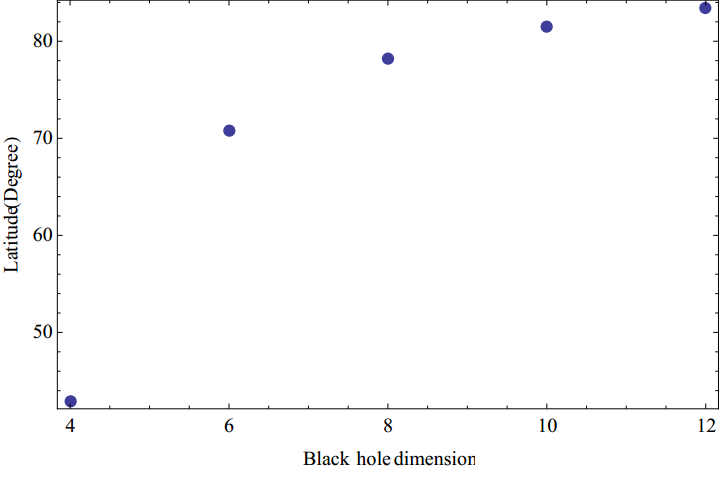}
	\caption{\label{lateven}Maximum amount of latitude is depicted for 4, ,6, 8, 10, and 12 dimensions.}
\end{figure}

In order to find the $\theta$ range in which particles can reach  arbitrary high energy, we need to use second derivative of effective potential which is defined in Eq.(\ref{b32}) and $\theta$ components of  geodesic particle which can be read from Eq.(\ref{b25}). By considering negative sign for the second derivative of effective potential $(V''(r)\leq 0)$, we can put an upper bound on constant $K$,  which is appear in the effective potential in Eq.(\ref{b31}). Also $\Theta$ function, which is defined in Eq.(\ref{b30}), should be greater than zero so that we can restrict $K$ from below. Finally, we can write a condition for $K$  as below
\begin{eqnarray}\label{d5}
	a^2\cos ^2\theta(m^2-E^2)+\frac{J_1^2}{\sin ^2\theta}\leq K \leq (\frac{2n+3}{2n-1}E^2-m^2)r_+^2+\frac{2n-1}{2n}J_1^2.
\end{eqnarray}
If we consider the critical particle has only one angular momentum, $J_1^2$ 
can be replaced by Eq.(\ref{d3}). Therefore, we can write inequality in Eq.(\ref{d5}) in the following form
\begin{eqnarray}\label{d6}
	&(m^2-E^2)(2n-1)\sin ^4\theta-[2nm^2-\frac{4(2n^3-2n^2+3n-1)}{(2n-1)^2}E^2]\sin ^2\theta
	\nonumber\\
	&-\frac{4n^2}{(2n-1)^3}E^2\geq 0.
\end{eqnarray}
Here, we consider a photon collides with another particle. In Fig.  \ref{lateven},  the maximum  amount of latitude is plotted for various even dimensional black holes. It is obvious that the area in which photon can reach arbitrary high energy is greater for higher  dimensional black holes.

\section{Efficiency of CM energy of two colliding particles}\label{sec5}
\begin{figure}[htbp!]
	\centering
	\includegraphics[scale=0.30]{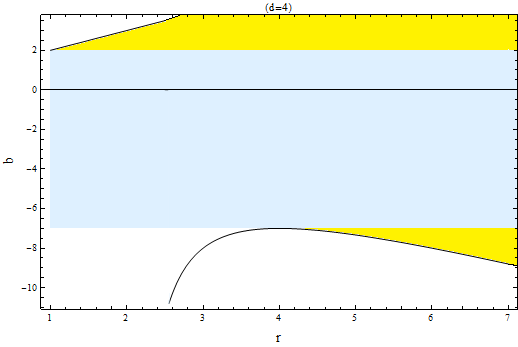}
	\hspace{0.5cm}
	\includegraphics[scale=0.20]{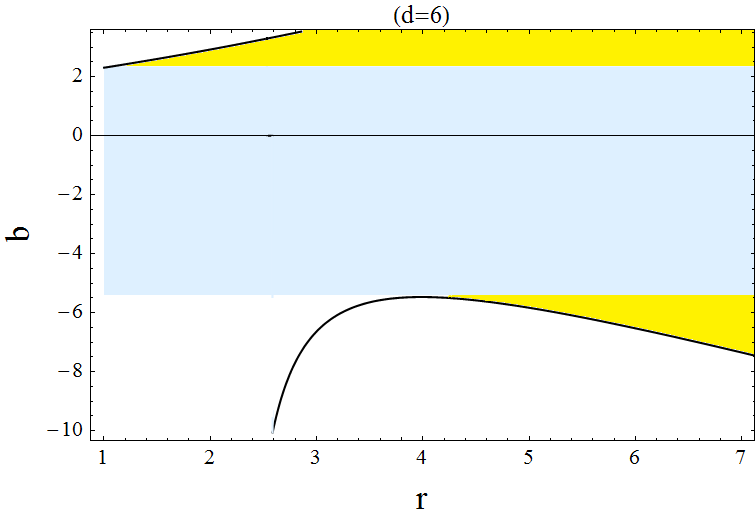}
	\caption{\label{impactparameter} Impact parameters of photons are depicted versus $r$ for 4 and 6 dimensions. Photon can escape to infinity in the yellow region. Event horizons of black holes are located in $r=1$.}
\end{figure}

\begin{figure}[htbp!]
	\centering
	\includegraphics[scale=0.25]{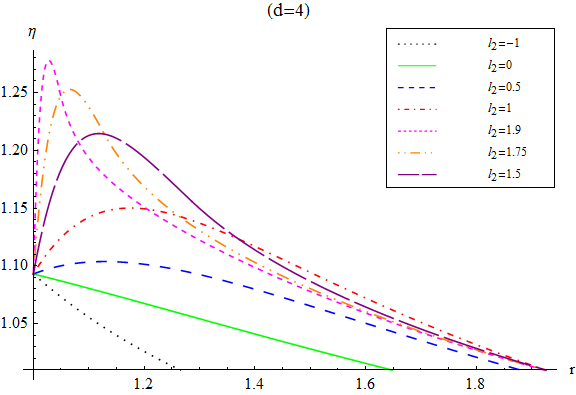}
	\hspace{0.5cm}
	\includegraphics[scale=0.25]{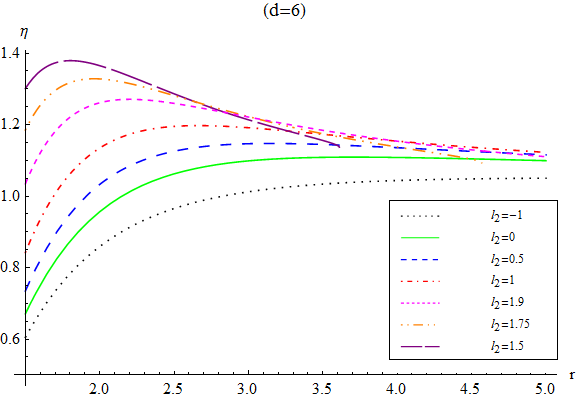}
	\caption{\label{randeman} In these plots, efficiency of two colliding particles are depicted versus r for different values of angular momenta in 4 and 6 dimensions. 
	}
\end{figure}

	In this Section, we investigate the efficiency of CM energy of a collision in the vicinity of MP black holes  \cite{12054350,14106446,151006764}. For simplicity, we just investigate a collision in which two massive particles of equal mass collide each other and produce two particles. 
We assume that  after collision one photon produced   that escapes to infinity 
and the other one falls into the black hole.
First, we need to study the impact parameter $(b=L/E)$ of the photon. In order to find the impact parameter, we should put effective potentials even dimensions black holes which are given in Eqs.(\ref{b31}) equal to zero and find their roots. The impact parameters of photons for 4 and 6 dimensions are depicted in Fig. {\ref{impactparameter}}. 
In order to find efficiency of collision, we consider two particles  collide and produce two particles that  $\bf p^{(3)}$ and $\bf p^{(4)}$ are producing particles   momenta and momenta of colliding particles are $\bf p^{(1)}$ and $\bf p^{(2)}$.
For simplicity, we assume that collision occur in equatorial plane of MP space-time. 
In four dimensional black holes,  six momentum components are unknown.
We consider conservation of energy 
\begin{eqnarray}
	E_{tot}\equiv E_1+E_2=E_3+E_4
\end{eqnarray}
and  conservation of angular an linear momenta
\begin{eqnarray}
	L_{tot}\equiv b_1E_1+b_2E_2=b_3E_3+b_4E_4\\
	p^{r}_{total}\equiv \epsilon _1P_1^r+\epsilon _2P_2^r=\epsilon _3P_3^r+\epsilon _4P_4^r
\end{eqnarray}
where $\epsilon _i$ can be equal to $+1$ or $-1$ for particle move inward or outward to black hole respectively. 
Using momenta  and energy conservation as constraints and the normalization condition ($p_\mu p_\nu g^{\mu\nu}=-m^2$), we can reach equation for $E_3$ which depends on angular momentum of intialling colliding particles, $E_3(b_3)$.  For higher dimensional black holes, we have more unknown  momentum components and more constraints, but in the end, one angular momentum will remain like the four-dimensional case.
By considering different amount for angular momentum of colliding particles, we can find maximum efficiency of collision  which is defined as $\eta=\frac{E_3}{E_1+E_2} $. 
In Fig. \ref{randeman}, efficiency of collisions for various amount of angular momenta of colliding particles are depicted for  4 and 6 dimensions.
Our results (Fig. \ref{randeman})  indicate
that maximum efficiency in 4 dimensions is smaller than 6 dimensions.
\section{Conclusion}\label{sec13}
	In this paper, a comprehensive analysis for colliding particles in higher dimensional black holes was presented.
The extremal black holes in higher dimensions can accelerate particles near their event horizons.
If the angular momenta of the particles are equal to the critical values the CM energy of two colliding particles blows up at the extremal event horizons of higher dimensional black holes.  
By using radial geodesic equation, we can define effective potentials.
The critical value of angular momenta make the effective potential and its first derivative zero.
In other word, critical value of angular momentum make radial velocity of particle zero.
Also, we  find a region in which two colliding particles can reach arbitrary high energy. 
For finding the region, we use second derivative of effective potential and the $\theta$ geodesic equations.
Finally , we calculate the efficiency of CM energy of two colliding particles.
Our basic result indicates that 
the higher dimensional black holes are better accelerators and more efficient  comparing to the lower dimensional ones.

\backmatter




\end{document}